\documentclass[a4paper,twocolumn,11pt, accepted=2022-08-03]{quantumarticle}
\pdfoutput=1
\usepackage[utf8]{inputenc}
\usepackage[english]{babel}
\usepackage[T1]{fontenc}
\usepackage{amsmath}
\usepackage{hyperref}

\usepackage{tikz}
\usepackage{lipsum}

\usepackage{graphicx}
\usepackage{dcolumn}
\usepackage{bm}
\usepackage{enumerate}
\usepackage{mathtools}
\usepackage{subcaption}
\usepackage{appendix}
\usepackage{braket}
\usepackage{physics}
\usepackage[linesnumbered,ruled,vlined]{algorithm2e}
\DeclareMathOperator*{\argmax}{arg\,max}
\usepackage{xcolor}
\usepackage{algpseudocode}
\usepackage{varwidth}

\begin{document}

\title{Quantum Local Search with the Quantum Alternating Operator Ansatz}

\author{Teague Tomesh}
\email{ttomesh@princeton.edu}
\affiliation{%
Department of Computer Science, Princeton University, Princeton, NJ 08540, USA.
}
\orcid{0000-0003-2610-8661}

\author{Zain H. Saleem}%
\email{zsaleem@anl.gov}
\affiliation{%
Argonne National Laboratory, 9700 S. Cass Ave., Lemont, IL 60439, USA.
}%

\author{Martin Suchara}%
\email{msuchara@anl.gov}
\affiliation{%
Argonne National Laboratory, 9700 S. Cass Ave., Lemont, IL 60439, USA.
}

\maketitle

\begin{abstract}
  We present a new hybrid, local search algorithm for quantum approximate optimization of constrained combinatorial optimization problems. We focus on the Maximum Independent Set problem and demonstrate the ability of quantum local search to solve large problem instances on quantum devices with few qubits. This hybrid algorithm iteratively finds independent sets over carefully constructed neighborhoods and combines these solutions to obtain a global solution. We study the performance of this algorithm on 3-regular, Community, and Erd\H{o}s-R\'{e}nyi graphs with up to 100 nodes.
\end{abstract}

\section{Introduction}
\label{sec:introduction}
The Quantum Approximate Optimization Algorithm (QAOA) \cite{farhi2014quantum} is a hybrid quantum-classical algorithm for finding the approximate solution to combinatorial optimization problems. This hybrid approach first encodes the problem's objective function as a Hamiltonian whose ground state corresponds to the optimal solution. Then the classical and quantum processors work together within a variational loop to find the ground state. The classical computer runs an optimization algorithm which traverses the optimization landscape searching for the extrema. During the course of the optimization the quantum processor is used to evaluate the expectation value of the objective function.

For unconstrained combinatorial optimization problems the optimization is performed over the entire Hilbert space generated by the variational ansatz (i.e., a parameterized quantum circuit). A new ansatz, the Quantum Alternating Operator Ansatz (QAO-Ansatz), was proposed in \cite{hadfield2018quantum, hadfield2019quantum} for solving \textit{constrained} combinatorial optimization problems. This variational ansatz is designed in such a way that the constraints are satisfied at all times and the optimization is performed only over the space of feasible solutions. 


Quantum Local Search (QLS) utilizes the QAO-Ansatz to find independent sets within small subgraphs (neighborhoods) whose size matches the capabilities of the quantum hardware. One of the main building blocks of the QAO-Ansatz is the mixing unitary which is defined up to a permutation of its components. QLS exploits this permutation freedom to search for optimal solutions within a neighborhood.
The QLS algorithm draws on methods from classical local search~\cite{aarts2003local} which are useful for problems where computing the global solution is intractable, but are amenable to decomposition into tractable subproblems. QLS constructs a global solution by iterating through many local subproblems and involves a dynamical update of the variational ansatz such that a constant amount of quantum resources are utilized. Local search strategies relying on graph partitioning have been previously applied in the quantum context, using the unconstrained variant of the QAOA, to optimization problems such as network community detection~\cite{shaydulin2019network}. However, the algorithmic components introduced in this work: combining local search techniques (neighborhood initialization and ansatz construction) with the constraint-preserving QAO-Ansatz have not been previously studied.

The finite size of quantum processors necessitates the development of algorithms, such as local search, which expand the applicability of hardware to larger problems sizes. As is the case with classical computers, even when we have a large-scale fault tolerant quantum computer in hand we will wish to solve problems beyond their current capacity. In this work, we use QLS to find approximate solutions over local neighborhoods within a larger graph and combine these into a final, global solution. While a quantum speedup obtained via quantum approximate optimization remains elusive, there is evidence that classically sampling from the output distribution of the QAOA is a difficult task~\cite{farhi2016quantum}. As the development of quantum optimization algorithms and quantum hardware continues to progress, these advancements may be incorporated within local search algorithms to continually expand the applicability of quantum computers. 

We study the performance of the QLS algorithm on the Maximum Independent Set (MIS) problem~\cite{karp1972reducibility}.
MIS is one of the most widely studied constrained combinatorial optimization problems, in part, due to its broad applicability in a variety of domains and the fact that it is equivalent to other important problems such as minimum vertex cover and maximum clique on its complement graph. Specifically, our contributions include:
\begin{enumerate}
    \item We introduce a method for constructing quantum circuits within a local neighborhood of a larger graph that are tunable to the size of the quantum hardware that is available. This tunability allows QLS to target graphs containing many more nodes than the number of qubits available in a particular quantum computer.
    \item We simulate and analyze the execution of the QLS algorithm on 3-regular, Community, and Erd\H{o}s-R\'{e}nyi graphs --- finding larger independent sets than those obtained with other classical and quantum algorithms.
    \item All of the code implementing the algorithms considered in this work has been made publicly available online~\cite{tomesh2021github} and is also available upon request.
\end{enumerate}

The remainder of the paper is organized as follows. In Section~\ref{sec:QAOA} we review the quantum approximate optimization algorithm and cover prior strategies for quantum constrained optimization. We introduce the quantum local search algorithm, provide its pseudocode, and an open source implementation~\cite{tomesh2021github} in Section~\ref{sec:qls}. Section~\ref{sec:results} provides the simulation results which compare the performance of QLS with other classical and quantum approaches. We study the impact of the free parameters within the QLS algorithm on runtime, performance, and quantum resources. Section~\ref{sec:conclusions} concludes and suggests future directions of this work. 
 
\section{Background}\label{sec:QAOA}
\subsection{Quantum Approximate Optimization}
Hybrid variational algorithms, like the QAOA~\cite{farhi2014quantum}, solve optimization problems by iteratively searching through the solution space with the combined efforts of a classical and quantum computer. The classical processor runs an optimization routine and calls the quantum processor to evaluate the computationally difficult objective function. For combinatorial optimization problems such as MIS, the problem is defined on a graph with $n$ vertices, and the graph-dependent classical objective function $C(\textbf{b})$ which we are looking to optimize is defined on $n$-bit strings $\textbf{b} = \{b_1,b_2,b_3 \dots b_n\} \in \{0,1\}^n$. The classical objective function can be written as a quantum operator diagonal in the computational basis:

\begin{equation}\label{qoperator}
   C_{obj} | b \rangle = C(\textbf{b}) |b \rangle .
\end{equation}

The expectation value of this objective function is measured with respect to the variational state,
\begin{equation}
|\psi_p(\boldsymbol{\gamma},\boldsymbol{\beta} )\rangle = e^{-i\beta_p M}e^{-i\gamma_p C}\dots e^{-i\beta_1 M}e^{-i\gamma_1 C}|s\rangle ,   
\label{eqn:ansatz}
\end{equation}
where $|s\rangle$ is the state on which we act with unitary operators to build our variational ansatz. The ansatz in Eq.~\ref{eqn:ansatz} is composed of two repeating parts: the phase separator unitary $e^{i \gamma_i C}$ and the mixing unitary $e^{i \beta_i M}$. The phase separator is a diagonal operator in the computational basis and typically takes the same form as the objective operator. The mixers are used to increase or decrease the amplitudes of different states -- effectively ``mixing'' the state of the current wavefunction. The variational parameters $\bm{\gamma}$ and $\bm{\beta}$ define the optimization landscape and correspond to the rotation angles of quantum gates within the ansatz.

For any variational state, the expectation value of $C_{obj}$
\begin{equation}
E_p(\boldsymbol{\gamma},\boldsymbol{\beta})=\langle  \psi_p(\boldsymbol{\gamma},\boldsymbol{\beta} )|C_{obj}|\psi_p(\boldsymbol{\gamma},\boldsymbol{\beta} )\rangle , 
\end{equation}
is evaluated on a quantum computer and then passed to a classical optimizer which attempts to find the optimal parameters that maximize $E_p(\boldsymbol{\gamma},\boldsymbol{\beta})$. Since the eigenstates of $C_{obj}$ are computational basis states, this maximization is achieved for the states corresponding to the solutions of the original combinatorial optimization problem. 

\subsection{Constrained Optimization: Maximum Independent Set}
When applying variational algorithms to unconstrained optimization problems, every basis state is a valid solution and therefore the optimization takes place over the entire Hilbert space generated by the variational ansatz. In contrast, constrained optimization is restricted to those basis states which satisfy the problem specific requirements. Hybrid variational algorithms have been adapted to constrained optimization problems in two main ways. Either the objective function is modified to heavily penalize invalid basis states, effectively turning the constrained problem into an unconstrained one~\cite{farhi2020quantum, farhi2020quantumA}, or the variational ansatz is structured in a way that keeps the optimization within the valid subspace~\cite{hadfield2018quantum, hadfield2019quantum, saleem2020max}.

Maximum Independent Set (MIS) is an NP-Hard constrained combinatorial optimization problem defined on the graph $G=(V,E)$ with nodes $V$, edges $E$ and number of nodes $n=|V|$ ~\cite{karp1972reducibility}. An independent set is defined as a subset $V' \subset V$ of the graph's nodes such that no two vertices in $V'$ share an edge. The goal of MIS is to find the independent set containing the largest number of nodes.

The Quantum Alternating Operator Ansatz (QAO-Ansatz)~\cite{hadfield2018quantum, hadfield2019quantum} is an example of an ansatz which imposes constraints at the quantum circuit level. The ansatz is constructed in such a way that we never leave the set of feasible states during the variational optimization (e.g. the set of all valid independent sets). For the MIS problem the objective function is the Hamming weight operator,
\begin{equation}
    C_{obj}= H = \sum_{i \in V} b_i
\end{equation}
where $b_i = \frac{1}{2} \left(1- Z_i\right)$ and $Z_i$ is the Pauli-Z operator acting on the $i$-th qubit. Each vertex in the graph is assigned a value $b_i \in \{0,1\}$ indicating whether it is excluded (0) or included (1) in the independent set. 

\begin{figure}[t!]
    \centering
    \includegraphics[width=\columnwidth]{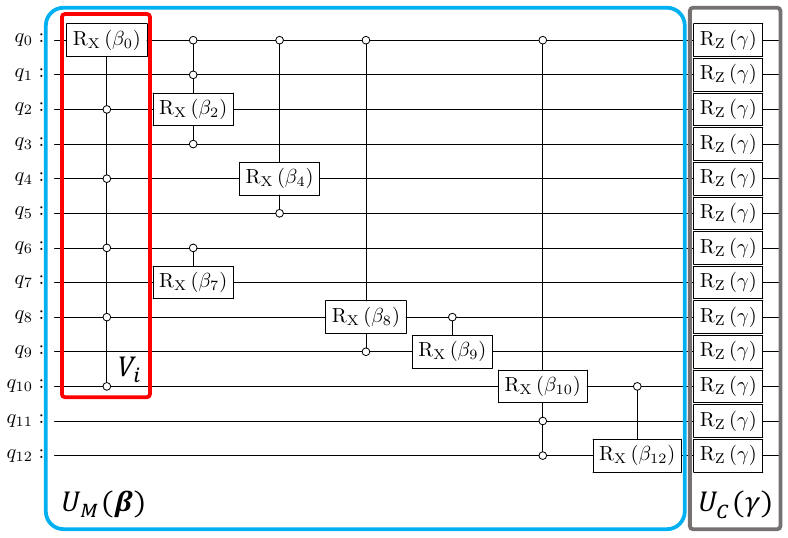}
    \caption{An example QLS ansatz containing: the mixer unitary (blue), partial mixer (red), and phase separator unitary (grey), based on the local neighborhood $G_{local}^0$ shown in Fig.~\ref{fig:examplegraph}.} 
    \label{fig:examplecircuit}
\end{figure}

The initial state $\ket{s}$ for the variational optimization can be any feasible state or superposition of feasible states. Similar to the unconstrained QAOA, the phase separator unitary for the QAO-Ansatz, $U_{C}(\gamma) \coloneqq e^{i \gamma H}$, is constructed using the objective function. However, the mixing unitary $U_{M}(\beta)$ is non-trivial and requires multi-qubit gates for its execution.
\begin{equation}
 U_{M}(\beta) \coloneqq \prod_j e^{i \beta M_j}\;,\;\;\;\; M_j = X_j \bar{B}
\end{equation}
and we have defined,
\begin{equation}
\bar{B} \coloneqq  \prod_{j=1}^{\ell} \bar{b}_{v_j}, \;\;\;\; \bar{b}_{v_j}=\frac{1+Z_{v_j}}{2},
\end{equation}
where $v_j$ are the neighbors and $\ell$ is the number of neighbors for the $i$th node. We can also write our mixer as
\begin{eqnarray}\label{eqn:mixer unitary}
\begin{aligned}
 U_M(\beta) & \coloneqq \prod_{j=1}^n e^{i \beta M_j} \\
  &= \prod_{j=1}^{n} \left(I + ( e^{-i\beta X_j}-I) \;  \bar{B} \right) \\ & \coloneqq \prod_{j=1}^{n}V_j(\beta) ,  
\end{aligned}
\end{eqnarray}
where we have used $\bar{b}_{v_j}^2=\bar{b}_{v_j}$.
The unitary mixer above is a product of $n$ partial mixers $V_i$, in general not all of which commute with each other $[V_i , V_j] \neq 0 $. 
The partial mixers in Eq.~\ref{eqn:mixer unitary} constitute multi-controlled X-rotations which can be implemented using multi-controlled Toffoli gates and controlled-X rotations. Examples of the phase separator, mixing, and partial mixer unitaries are shown in Fig.~\ref{fig:examplecircuit}. The effect of applying a partial mixer $V_i(\beta)$ can be stated in words as: if all of node $i$'s neighbors are in the $\ket{0}$ state (i.e. are not included in the current independent set), then rotate qubit $i$'s state around the X-axis by an angle $\beta$.


\section{Quantum Local Search}\label{sec:qls}

\begin{figure}[t!]
    \centering
    \includegraphics[width=\columnwidth]{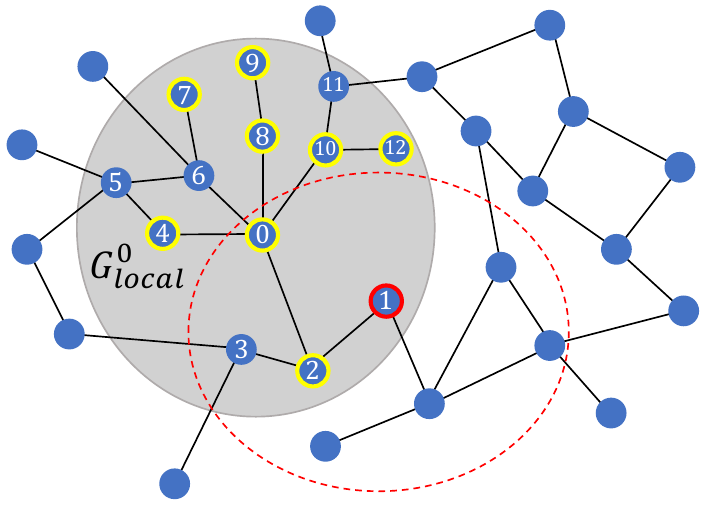}
    \caption{QLS example: the neighborhood (grey circle) $G_{local}^0$ of distance $N_s=2$ surrounds the root node $n_0$. The yellow highlighted nodes indicate where the partial mixers may be applied since all of their neighbors are contained within $G_{local}^0$. The nodes in $G_{local}^0$ are labelled to match the circuit shown in Fig.~\ref{fig:examplecircuit}. After the variational optimization is complete, a new root node, $n_1$ is selected and will induce a new subgraph (red dotted circle). The size of the neighborhood $N_s$ can be scaled to match the problem instance or the available quantum resources.} 
    \label{fig:examplegraph}
\end{figure}

\begin{algorithm}[t!]
\caption{Quantum Local Search}
 \label{alg:qls}
\SetAlgoLined
\SetKwData{Left}{left}\SetKwData{This}{this}\SetKwData{Up}{up}
\SetKwFunction{Union}{Union}\SetKwFunction{FindCompress}{FindCompress}
\SetKwInOut{Input}{Input}\SetKwInOut{Output}{Output}

\begin{varwidth}[t]{\linewidth}
    \textbf{Input:} $G=(V,E)$, $r$: permutation rounds,\par
    \hskip \algorithmicindent $N_{pm}$: partial mixers,\par
    \hskip \algorithmicindent $N_s:$ neighborhood size
\end{varwidth}

\Output{$S$: Approximate MIS of $G$}
$S \leftarrow [0, 0, \dots, 0]$, length $|V|$ bitstring\;
\Repeat{\text{all nodes have been visited}}{
  \tcc{Neighborhood\\ Initialization (Sec.~\ref{subsec:one})}
  $n_{root} \leftarrow$ select a root node (Sec.~\ref{subsec:four})\;
  $G_{local} \leftarrow$ subgraph of $n_{root}$ up to distance $N_s$\;
  $s \leftarrow \bigotimes_{i \in \text{nodes}(G_{local})} \ket{S_i}$ \;
  \tcc{Neighborhood Ansatz \\ Construction (Sec.~\ref{subsec:two})}
  $\bm{\beta} \leftarrow [0_1, 0_2, \dots, 0_m]$\;
  $\gamma \leftarrow$ random$(0,2\pi)$\;
  \tcc{Randomly initialize $N_{pm}$ \\ partial mixers starting \\ from $n_{root}$}
  \For{$i \in [N_{pm}]$}{
    $\beta_i \leftarrow$ random$(0,2\pi)$\;
  }
  $U_{qls} \leftarrow U_M(\bm{\beta}) U_C(\gamma)$\;
  
  \tcc{Neighborhood Solution\\ Search (Sec.~\ref{subsec:three})}
  bitstrs $\leftarrow [\; ]$\;
  \For{$i \in r$}{
    $U_M(\bm{\beta}) \leftarrow \mathcal{P}\big(V_1(\beta_1)V_2(\beta_2)\cdots V_m(\beta_m)\big)$\;
    \While{not converged}{
      counts $\leftarrow$ execute($U_{qls}(\bm{\beta}, \gamma)\ket{s}$)\;
      $E \leftarrow$\ expectation\_value($H$, counts)\;
      $\bm{\beta}, \gamma \leftarrow \text{update\_params(E)}$\;
    }
    
    \begin{varwidth}[t]{\linewidth}
      bitstrs.append$($\par
      \hskip \algorithmicindent $\argmax_b{([H(b) \text{ for } b \in \text{counts}]))}$\;
    \end{varwidth}
  }
  \tcc{Update global solution (Sec.~\ref{subsec:five})}
  \begin{varwidth}[t]{\linewidth}
      $update(S,$\par
      \hskip \algorithmicindent $\argmax_b{([H(b) \text{ for } b \in \text{bitstrs}])})$\;
  \end{varwidth}
  
}
 \KwRet{S}
 
\end{algorithm}

The Quantum Local Search (QLS) algorithm finds approximate solutions to the MIS problem on a graph $G=(V,E)$ with $n$ vertices by iteratively optimizing a variational ansatz over small neighborhoods within $G$. 
We give an outline of the QLS pseudocode in Alg.~\ref{alg:qls} and an implementation is available via Github~\cite{tomesh2021github}.

\subsection{Neighborhood Initialization}\label{subsec:one}
Prior to the start of the algorithm, initialize a solution bitstring $S = \{0\}^n$ to the all-zero state with length $n$; this will store the current global approximate solution and be updated throughout the course of the algorithm. Select a root node $n_{0}$ and its corresponding local subgraph $G^0_{local} = (V_{local}, E_{local})$ where all the nodes in this subgraph are a node distance $N_s$ away from $n_{0}$ (see Fig.~\ref{fig:examplegraph} for an example). The distance $N_s$ is a free parameter used to set the size of the neighborhood. This parameter should be set according to the density of the target graph such that the number of nodes $m$ in the neighborhood does not exceed the number of qubits available in the quantum hardware. 
In this work, the qubits in the initial state $\ket{s}$ of the neighborhood are initialized to match their current state within $S$, explicitly $\ket{s} = \bigotimes_{i \in V_{local}} \ket{S_i}$. However, more interesting states could also be used while respecting the MIS constraint~\cite{egger2021warm, saleem2021approaches}.

\subsection{Neighborhood Ansatz Construction}\label{subsec:two}
For simplicity, the example ansatz construction covered here will be restricted to the case with depth parameter $p=1$, but this method can be readily extended to $p>1$.
For depth $p=1$ the variational ansatz $U_{qls}(\bm{\beta}, \gamma) = U_M(\bm{\beta}) U_C(\gamma)$ is composed of a single layer of the phase separator and mixing unitaries  and produces a state
\begin{equation}\label{eqn:qls ansatz}
    |\psi (\bm{\beta}, \gamma)\rangle  = U_{qls}(\bm{\beta}, \gamma)\ket{s} = U_M(\bm{\beta}) U_C(\gamma) \ket{s}.
\end{equation}
The phase separator $U_C(\gamma) = e^{i\gamma H}$ is identical to that used in the QAO-Ansatz.
The mixer unitary is parameterized by a set of $m$ angles $\bm{\beta} = (\beta_1, \beta_2, ..., \beta_m)$ corresponding to the partial mixers of each node in the subgraph.
\begin{equation}
    U_M(\bm{\beta}) = \prod_{i=1}^{m} V_i(\beta_i) = \prod_{i=1}^{m} \left(I + ( e^{-i\beta_i X_i}-I) \;  \bar{B} \right),
\end{equation}
Prior work, investigating the application of the QAOA to MaxCut problems, has shown that allowing the partial mixers to be uniquely parameterized in this way can lead to increased approximation ratio and reduced circuit depth at the expense of an increased number of variational parameters~\cite{herrman2022multi}. However, the QLS ansatz differs from that considered in \cite{herrman2022multi} because the partial mixers do not commute with one another. This allows us to randomize over different permutations of the partial mixers to escape local minima.

Applying all $m$ partial mixers at once may require an intractable amount of quantum resources, especially for larger neighborhoods. Instead, we use a hyperparameter $N_{pm}$ which sets the number of nonzero $\beta_i$ within $\bm{\beta}$. With $N_{pm}$ set, we start by applying the partial mixers first to the central node $n_0$, then to the nodes that are distance one away from the central node and then distance two and so on until we have either reached the quota of $N_{pm}$ partial mixers or exhausted the nodes in the neighborhood. Fig.~\ref{fig:examplecircuit} shows an example QLS ansatz corresponding to the neighborhood constructed in Fig.~\ref{fig:examplegraph}.

In the process of constructing $U_{qls}$ it is possible that we could find a high degree node within the neighborhood which has more neighbors than there are qubits available in the quantum hardware (or equivalently, exceeds the amount of allocated resources). We can handle this corner case by simply skipping the application of this node's partial mixer. This node can still participate in its neighbor's partial mixers as a control qubit. 

\subsection{Neighborhood Solution Search}\label{subsec:three}
Once the circuit construction is finished, we run the quantum approximate optimization algorithm with the goal of finding the parameters $\bm{\beta}$ and $\gamma$ in Eq.~\ref{eqn:qls ansatz} that maximize

\begin{equation}
  \bra{\psi(\bm{\beta}, \gamma)} H \ket{\psi(\bm{\beta}, \gamma)}.
\end{equation}

This will output a bitstring with a certain Hamming weight.
Since the mixer unitary is defined up to a permutation of the partial mixers:

\begin{equation}
U_M(\bm{\beta}) \simeq \mathcal{P}\big(V_1(\beta_1)V_2(\beta_2)\cdots V_m(\beta_m)\big), 
\end{equation}

different permutations can return bitstrings with different Hamming weights.
We can rerun this step $r$ number of times, each time randomly choosing a different permutation of the partial mixers. This is a very useful strategy for escaping local minima which is only possible due to the definition of $U_M(\bm{\beta})$ up to a permutation --- something which is not available to the typical implementation of the QAOA. From these $r$ different rounds we select the bitstring with the largest Hamming weight, and update the solution bitstring $S$ with the newly found local solution. This step is entirely parallelizable and multiple QPUs may be employed to simultaneously explore different permutations of the partial mixers.

\subsection{Neighborhood Update}\label{subsec:four}
Once we have obtained an independent set on the neighborhood $G^0_{local}$ we traverse the graph $G$ by selecting a new root node.
The root node $n_1$, red highlighted node in Fig.~\ref{fig:examplegraph}, of the next neighborhood $G^1_{local}$, red dashed circle in Fig.~\ref{fig:examplegraph}, is randomly selected from the set of vertices that are on the edge of the current neighborhood, a distance $N_s$ away from $n_0$. 

\subsection{Obtaining an Approximate MIS on $G$}\label{subsec:five}
We then repeat steps 1 through 4 starting with the new root node, and continue this process until all nodes have participated in the local search. Throughout the entire execution the bitstring $S$ is continually updated with the solutions found on $G^i_{local}$ and a global approximate solution over the full graph $G$. Once the entire graph has been traversed the bitstring 
\begin{equation} \nonumber
    S = \{b_i\}^n, \quad b_i = 
    \begin{cases}
        0 & V_i \notin \mathcal{I}\\
        1 & V_i \in \mathcal{I}
    \end{cases}
\end{equation}
 is returned representing the final independent set $\mathcal{I}$.

\section{Experimental Evaluation}\label{sec:results}

\subsection{Methodology}
\begin{figure}
    \centering
     \begin{subfigure}[b]{0.3\columnwidth}
         \centering
         \includegraphics[width=\linewidth]{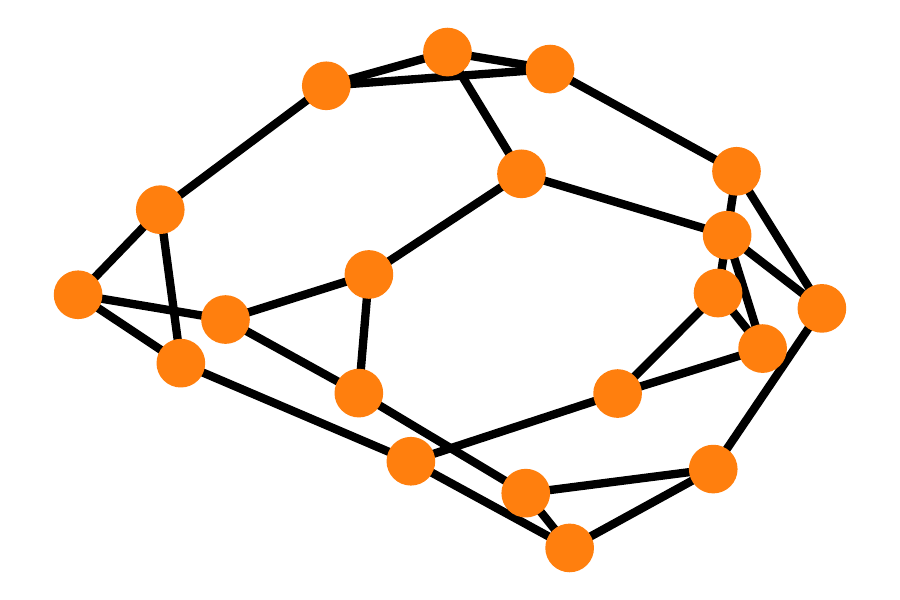}
         \caption{}
         \label{fig:regular}
     \end{subfigure}
    \hfill
     \begin{subfigure}[b]{0.3\columnwidth}
         \centering
         \includegraphics[width=\linewidth]{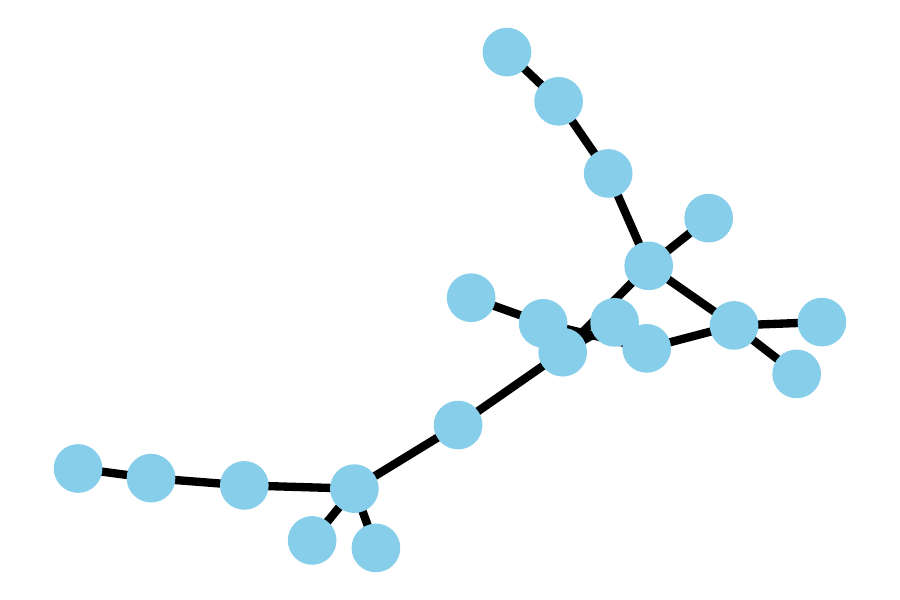}
         \caption{}
         \label{fig:community}
     \end{subfigure}
     \hfill
     \begin{subfigure}[b]{0.3\columnwidth}
         \centering
         \includegraphics[width=\linewidth]{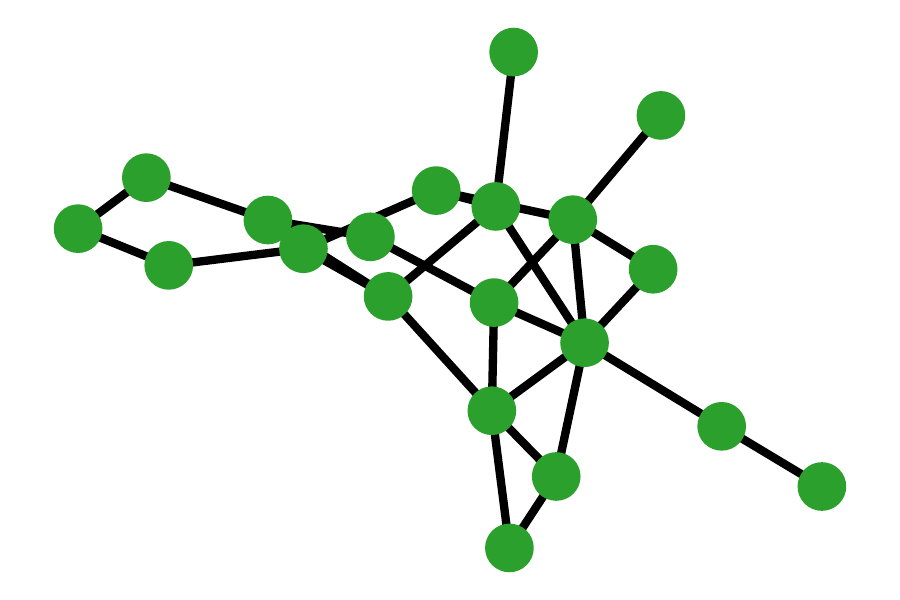}
         \caption{}
         \label{fig:erdosrenyi}
     \end{subfigure}
    \caption{Example (a) 3-regular, (b) (0.1, 0.02)-community, and (c) (12, 18)-Erd\H{o}s-R\'{e}nyi graphs used as benchmarks in this work.}
    \label{fig:examplegraphs}
\end{figure}

We benchmark the performance of QLS, as well as additional classical and quantum algorithms, on three different graph types containing 20, 60, and 100 vertices. We consider $d$-regular, ($p_{in}$, $p_{out}$)-Community, and ($n$,$m$)-Erd\H{o}s-R\'{e}nyi graphs. Examples of these are shown in Fig.~\ref{fig:examplegraphs}. We include the different graph types to obtain a clearer picture of algorithmic performance in a variety of circumstances. 

Prior work on the QAOA has often focused exclusively on 3-regular graphs because their regularity allows for easier analysis of asymptotic performance~\cite{farhi2014quantum}. However, in practice many graphs drawn from real-world data do not exhibit the rigid structure of $d$-regular graphs. For example, many social media or protein-protein interaction networks contain a community-like structure where subsets of nodes are densely connected with fewer edges existing between communities~\cite{fortunato2010community}. To capture this class of networks we include the community graphs generated by the \texttt{planted\_partition\_graph()} function within the NetworkX Python package~\cite{hagberg2008exploring}. This function takes as input the number and size of the communities as well as the probability of edges existing within and between communities. We consider (0.1, 0.02)-Community graphs with community sizes of 20 nodes and set the probability of an edge existing within a community to $p_{in}=0.1$ and between communities as $p_{out}=0.02$. Finally, we also include a class of random graphs known as Erd\H{o}s-R\'{e}nyi graphs which have been used in prior work to study the asymptotic performance of the QAOA on MIS problems~\cite{farhi2020quantumA}. We use the same construction as \cite{farhi2020quantumA} for average degree Erd\H{o}s-R\'{e}nyi graphs where the number of nodes $n$ and the average degree $d$ are fixed and exactly $\frac{dn}{2}$ edges are randomly chosen from a uniform distribution. We consider ($n$, $\frac{dn}{2}$)-Erd\H{o}s-R\'{e}nyi benchmark graphs with $d=3$.

\begin{figure}[t!]
    \centering
    \includegraphics[width=\columnwidth]{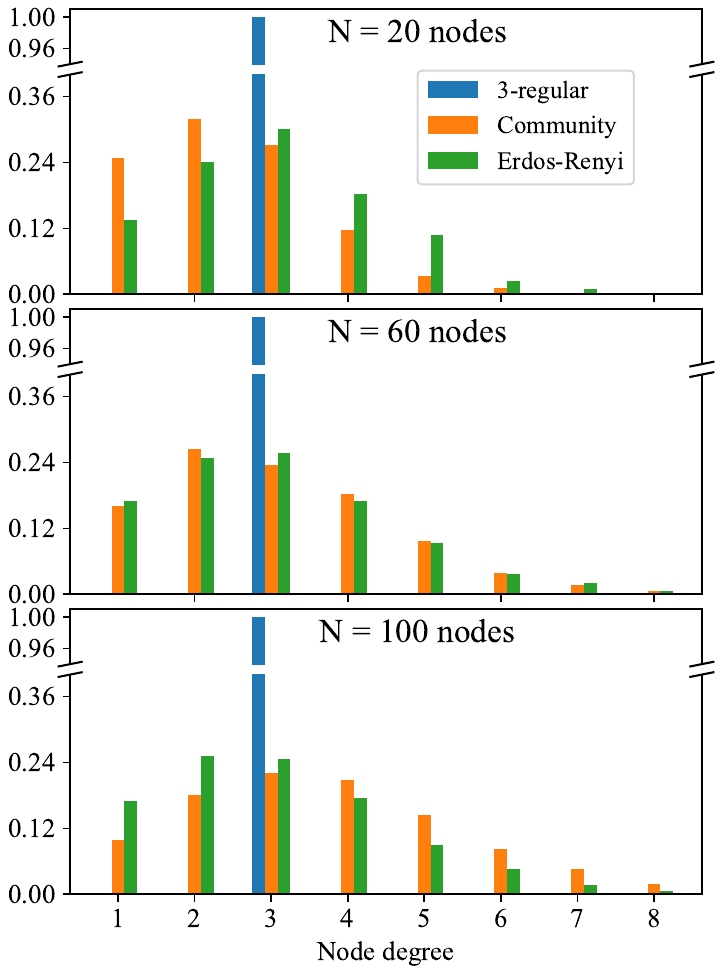}
    \caption{Distributions of node degree for $n=20,60,100$ node 3-regular, (0.1, 0.02)-Community, and ($n$, $\frac{3n}{2}$)-Erd\H{o}s-R\'{e}nyi graphs.}
    \label{fig:graphstats}
\end{figure}

\subsubsection{Keeping Simulations Tractable}

All simulations were performed using the Princeton Ionic cluster~\cite{ionic} with a memory budget of 10GB of RAM per job.
For each graph type and size $n={20,60,100}$ we generated 40 random instances to serve as the benchmarks. We chose the parameters of the community and Erd\H{o}s-R\'{e}nyi graphs such that each node has approximately three neighbors. This was done to ensure that the simulations of QLS remained tractable when increasing the size of the neighborhood and number of partial mixers utilized. By setting the graph parameters in this way, the circuits we encountered in our simulations never contained more than  25 qubits which kept the runtime and memory requirements manageable. However, despite this limitation, the benchmark graphs still display varying structure with a mix of high- and low-degree nodes as shown in Fig.~\ref{fig:graphstats}.

\subsubsection{Performance Metric}
The typical measure of performance for MIS is the approximation ratio between the independent set found by an algorithm and the optimal MIS. Since finding the optimal MIS for large graphs quickly becomes intractable, we instead measure performance using the independence ratio 
\begin{equation} \label{eqn:independenceratio}
    R = \frac{H(b)}{n}
\end{equation}
defined as the size of the independent set returned by the algorithm ($H(b)$ is the Hamming weight of the bitstring assignment of nodes in the independent set) divided by the size of the graph $n$~\cite{farhi2020quantumA}.

We compare the performance of a number of different algorithms for MIS including quantum local search (QLS), classical local search (CLS), Boppana-Halld\'{o}rsson~\cite{boppana1992approximating}, QAOA+~\cite{farhi2020quantumA, saleem2021approaches}, and greedy random search. The implementations of each of these algorithms is provided in the following section. To mitigate the effects that our choice of classical optimizer has on the performance of the QLS and QAOA+ algorithms we repeat the evaluation of \textit{all} algorithms five times on each benchmark graph and select the best run. These results are then averaged over all benchmark graphs.

\begin{figure*}[t]
    \centering
    \includegraphics[width=\textwidth]{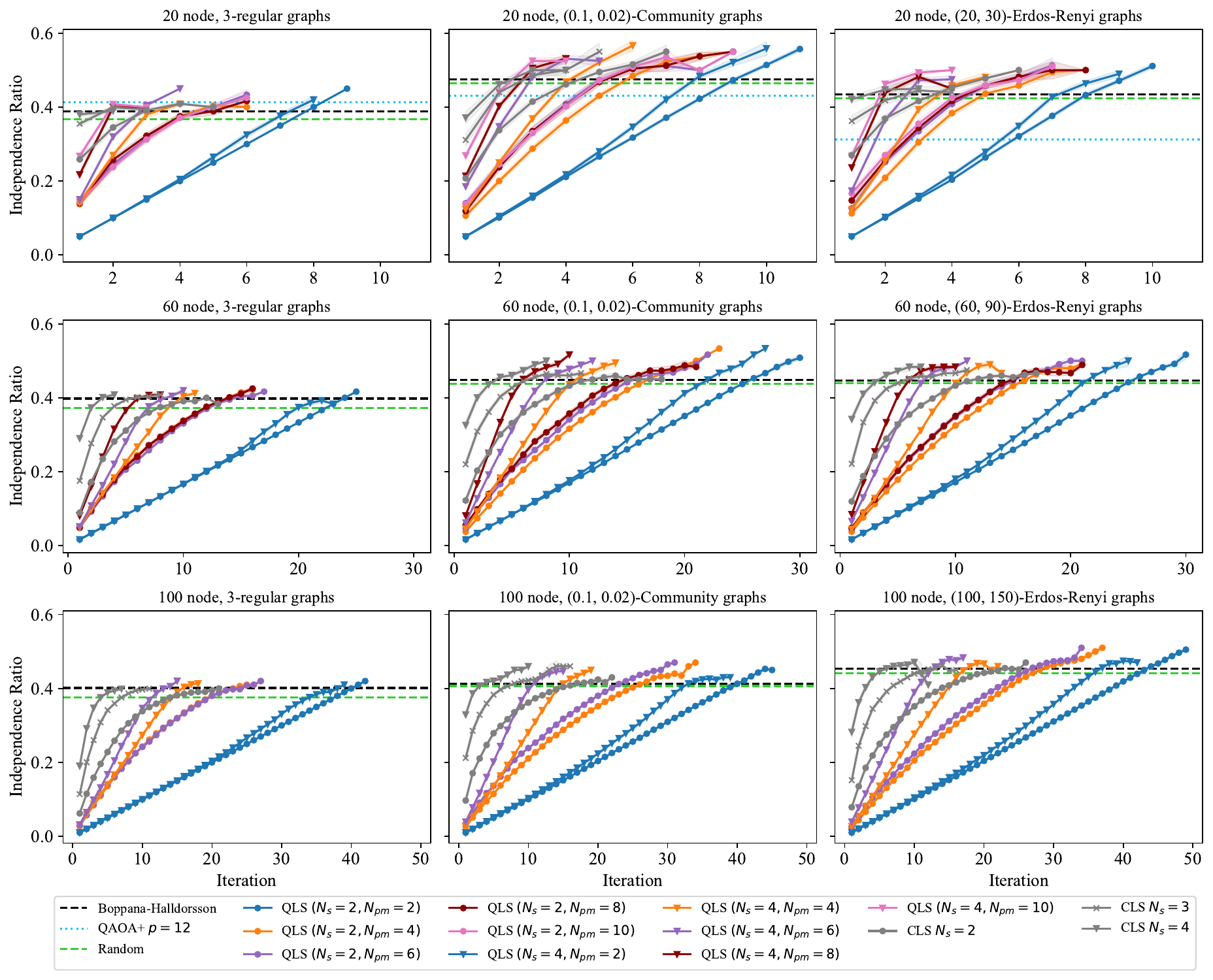}
    \caption{Average independence ratio (Eq.~\ref{eqn:independenceratio}) of the independent sets found by QLS and CLS at each iteration of the algorithm collected over 40 benchmark graphs. Shaded areas indicate the standard error from the mean. Both QLS and CLS converge to a solution faster as the number of partial mixers and neighborhood size is increased.}
    \label{fig:convergence}
\end{figure*}

\begin{figure*}[t]
    \centering
    \includegraphics[width=\textwidth]{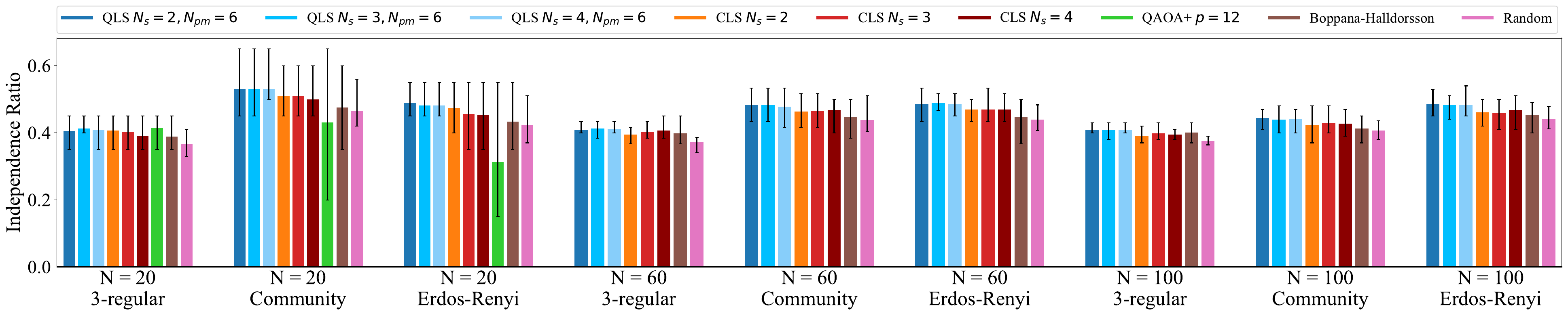}
    \caption{Average final independence ratio produced by the different algorithms considered in this work. The error bars denote the maximum and minimum independence ratios seen over the set of 40 benchmark graphs.}
    \label{fig:allbars}
\end{figure*}

\subsection{Implementation}
All of the code used to implement the MIS algorithms in this work have been made open source and are available online~\cite{tomesh2021github}. The implementation of the Boppana-Halld\'{o}rsson algorithm is available via the NetworkX package~\cite{hagberg2008exploring}. Note that this algorithm evaluates the entire problem graph at once and does not contain any iterative local search component. Additionally, as a baseline for performance comparisons we implemented a randomized algorithm for MIS which randomly visits every node in the graph and greedily adds nodes to the independent set whenever possible.

\subsubsection{Quantum Local Search}
The QLS specification described in Algorithm~\ref{alg:qls} was implemented in Python with Qiskit \texttt{v0.34.1} ~\cite{cross2018ibm}. The quantum circuit simulations were performed via statevector simulation using the Qiskit Aer submodule. For the simulation results shown in this work we set the number of permutation rounds $r=3$ and varied the neighborhood size $N_s={2,3,4}$ and the number of partial mixers $N_{pm}={2, 3, \dots, 10}$ while keeping the maximum circuit size below 25 qubits. This bound stems from our use of classical statevector simulation and the maximum $N_s$ and $N_{pm}$ values vary depending on the graph type. 

\subsubsection{Classical Local Search}
We implemented a classical analogue of QLS which utilizes the same neighborhood initialization step described in Sec.~\ref{sec:qls}. However, instead of finding an approximate MIS solution using a hybrid quantum-classical algorithm we apply the Boppana-Halld\'{o}rsson algorithm to the full neighborhood. This implementation of the Boppana-Halld\'{o}rsson algorithm has been slightly modified from the one provided in NetworkX to allow for the input of an initial independent set. This is useful in the local search context as the current neighborhood may contain nodes that are already within the independent set and so this information should be reflected in the Boppana-Halld\'{o}rsson execution.

Additionally, the CLS algorithm requires an additional step after obtaining an approximate solution over the neighborhood to ensure that no adjacent nodes along the edge of the neighborhood were both added to the independent set. This is not an issue within QLS because while a node on the edge of the neighborhood (meaning it has at least one neighbor outside of the local neighborhood) may participate as a control qubit it does not have it's own partial mixer applied. In the CLS case, after a local solution is obtained we iterate through each node added to the independent set and check whether any of its neighbors are also in the set. If so, we remove the current node from the independent set.

\subsubsection{QAOA+}
The QAOA was originally introduced within the context of unconstrained optimization problems such as MaxCut~\cite{farhi2014quantum}. A modified version, known as QAOA+, was later adapted to constrained problems such as MIS by incorporating the independence constraint within the objective function~\cite{farhi2020quantumA}. The additional term added to the objective function of QAOA+ penalizes bitstrings which are invalid independent sets; effectively turning the constrained optimization into an unconstrained one. The invalid bitstrings output by QAOA+ can be pruned in an additional post-processing step. The implementation of QAOA+ used in this work is based on that given in \cite{saleem2021approaches}.

\subsection{Results}

\begin{figure}[t]
    \centering
     \begin{subfigure}{0.49\columnwidth}
         \centering
         \includegraphics[width=\linewidth]{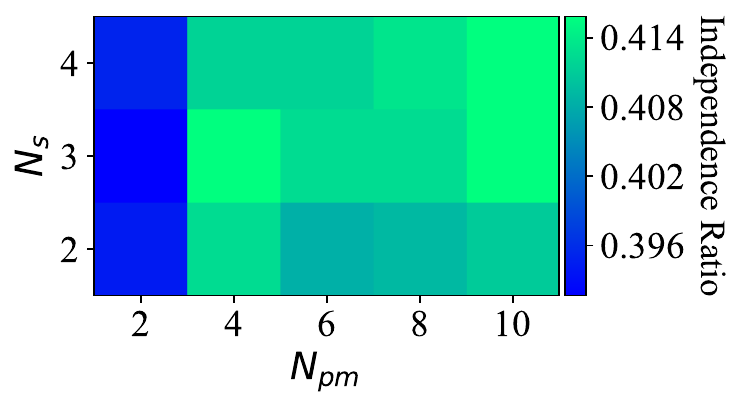}
         \caption{3-regular}
         \label{fig:heatmap_regular_performance}
     \end{subfigure}
    \hfill
     \begin{subfigure}{0.49\columnwidth}
         \centering
         \includegraphics[width=\linewidth]{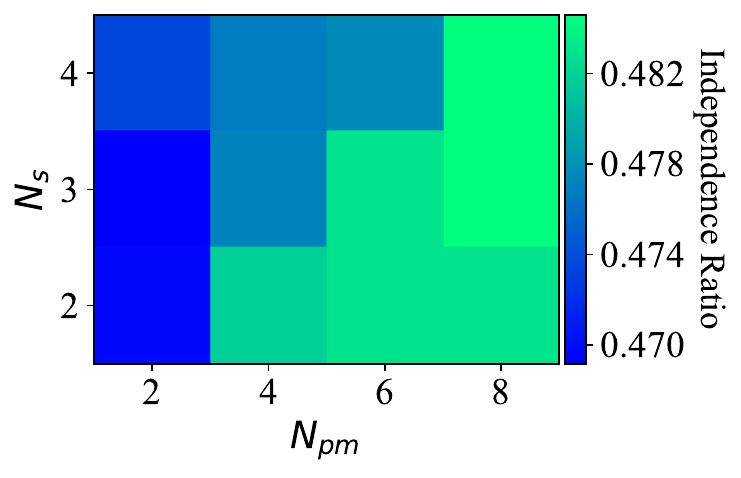}
         \caption{$(0.1, 0.02)$-Community}
         \label{fig:heatmap_community_performance}
     \end{subfigure}
     
     \medskip
     \begin{subfigure}{0.49\columnwidth}
         \centering
         \includegraphics[width=\linewidth]{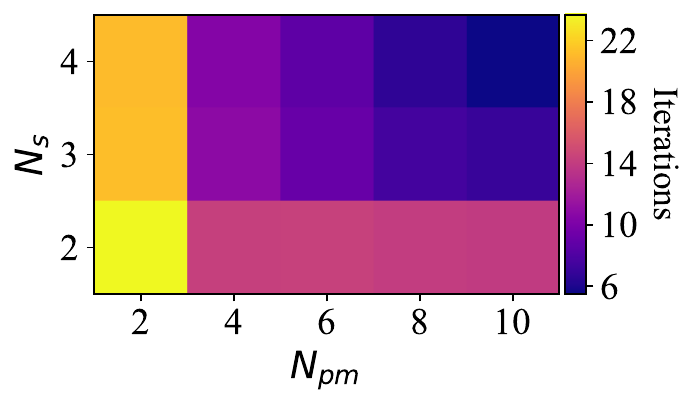}
         \caption{3-regular}
         \label{fig:heatmap_regular_runtime}
     \end{subfigure}
     \hfill
     \begin{subfigure}{0.49\columnwidth}
       \centering
       \includegraphics[width=\linewidth]{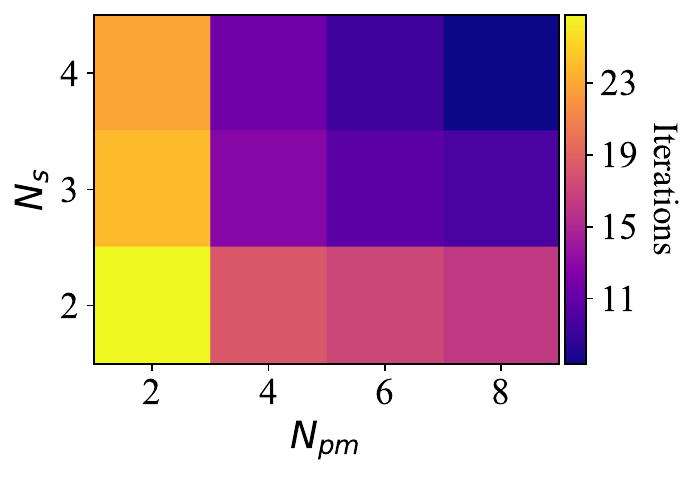}
       \caption{$(0.1, 0.02)$-Community}
       \label{fig:heatmap_community_runtime}
     \end{subfigure}
    \caption{Performance and runtime of QLS on $N=60$ node graphs with varying $(N_s, N_{pm})$ values.}
    \label{fig:heatmaps}
\end{figure}

\begin{figure*}[t]
    \centering
     \begin{subfigure}[b]{0.3\textwidth}
         \centering
         \includegraphics[width=\linewidth]{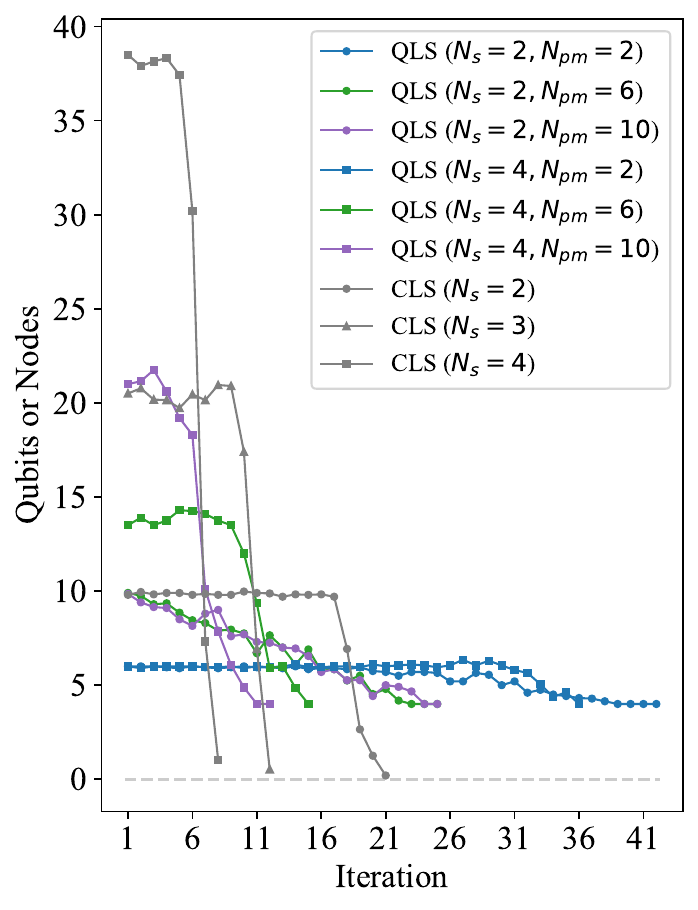}
         \caption{3-regular}
         \label{fig:d3sizes}
     \end{subfigure}
    \hfill
     \begin{subfigure}[b]{0.3\textwidth}
         \centering
         \includegraphics[width=\linewidth]{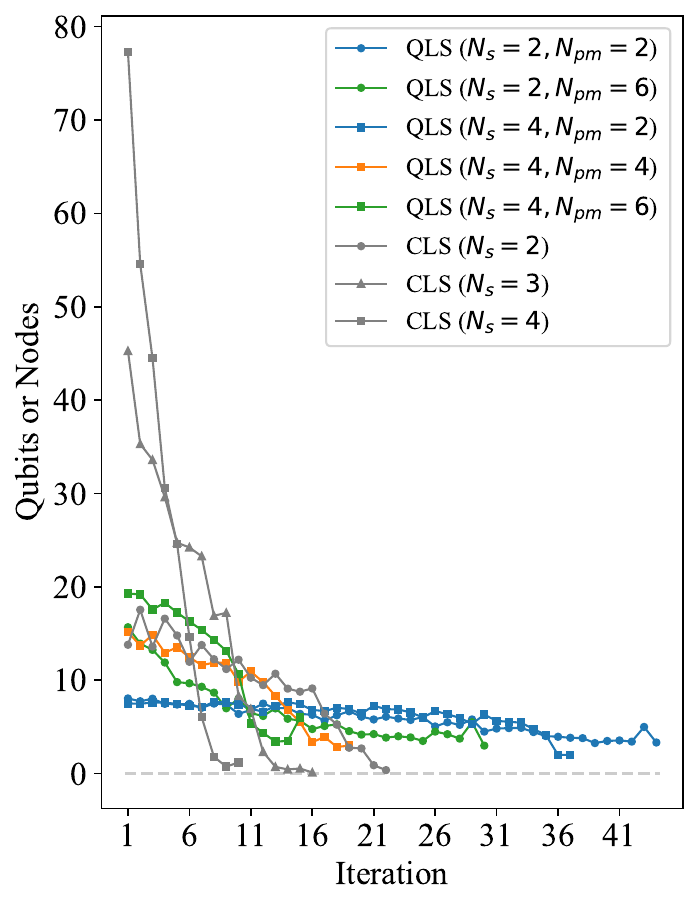}
         \caption{$(0.1, 0.02)$-Community}
         \label{fig:ppsizes}
     \end{subfigure}
     \hfill
     \begin{subfigure}[b]{0.3\textwidth}
         \centering
         \includegraphics[width=\linewidth]{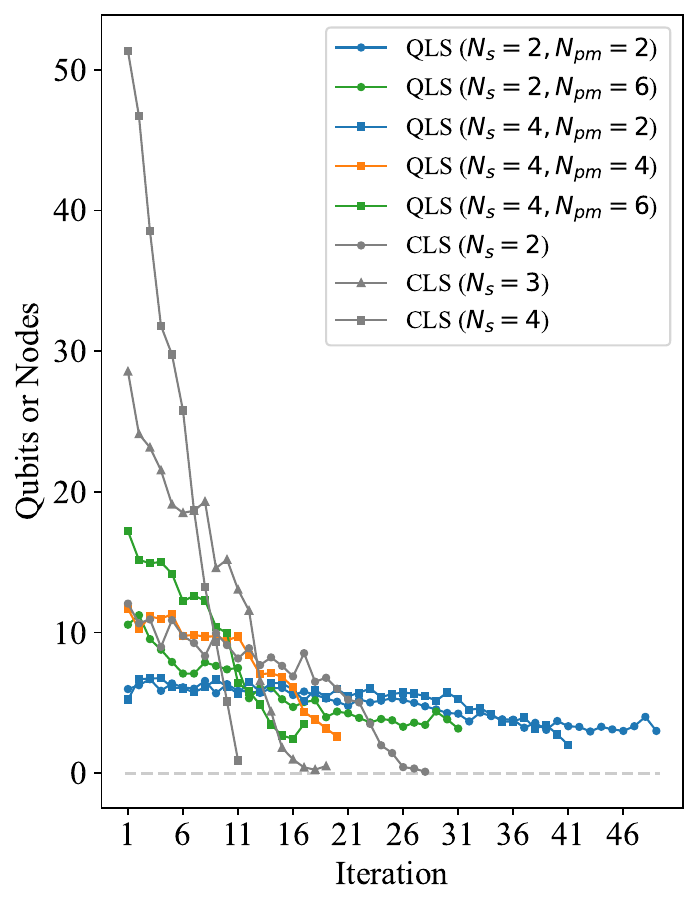}
         \caption{$(100, 150)$-Erd\H{o}s-R\'{e}nyi}
         \label{fig:ersizes}
     \end{subfigure}
    \caption{Average circuit size during each iteration of QLS on $N=100$ node graphs. While the CLS algorithm does not construct quantum circuits, the data here shows the average size of the local neighborhood at each iteration.}
    \label{fig:circuitsizes}
\end{figure*}

The performance of the quantum and classical algorithms over the course of the local search iterations is shown in Fig.~\ref{fig:convergence}. The data points indicate the average independence ratio across all benchmark graphs at the current iteration, and the shaded areas denote the standard error from the average value. Averaging the results in this way can result in sharp jumps at the final iterations, this is due to the fact that some executions may have terminated at an earlier iteration and therefore the average is performed over a smaller set of data. In Fig.~\ref{fig:allbars} we remove the dependence on iteration number and report the final average independence ratio to account for this visual artifact.

The QLS algorithm was evaluated for a range of $(N_s, N_{pm})$ parameter pairs, and their effect on the runtime can be seen in Fig.~\ref{fig:convergence}  (Fig.~\ref{fig:heatmaps} also shows this dependence for a subset of the data). The value of $N_{pm}$ controls the number of partial mixers allowed within the ansatz during each iteration of the algorithm. As more quantum resources are utilized more vertices can be added to independent set during a single iteration (indicated by the slope of the lines in Fig.~\ref{fig:convergence}). The rate at which the independent set grows remains fairly constant during the early iterations of the local search, but begins to decline as more of the nodes are visited until finally the program halts.

The neighborhood size $N_s$ can impact this execution in a couple of different ways. For lower values of $N_s$ the neighborhood may only support a limited number of partial mixers and increasing $N_{pm}$ beyond this threshold no longer improves runtime. This is seen clearly in the 20 node, 3-regular graph simulations in Fig.~\ref{fig:convergence} where for $N_s = 2$ the lines for $N_{pm} = 4, 6, 8,$ and $10$ all lie on top of one another, indicating that the local neighborhood is saturated with just 4 partial mixers. This effect also appears in Fig.~\ref{fig:heatmap_regular_runtime} where increases in $N_{pm}$ do not result in decreases in iteration count. 

Given a specific neighborhood size the exact threshold of partial mixers it can support depends on the graph type since higher degree nodes can yield more nodes within a single neighborhood. This is evident in the 20 node, Community and Erd\H{o}s-R\'{e}nyi data of Fig.~\ref{fig:convergence} where, unlike the 3-regular case, runtime improvements can still be found between the $(N_s=2, N_{pm}=4)$ and $(N_s=2, N_{pm}=6)$ evaluations. Furthermore, for graphs that contain higher degree nodes larger circuit sizes are encountered even when using fewer partial mixers since a single high-degree node will translate into a partial mixer with many control qubits. Fig.~\ref{fig:circuitsizes} demonstrates this effect where 20 qubit circuits only appear during the course of $(N_s=4, N_{pm}=10)$ QLS on 3-regular graphs, but are already present in $(N_s=4, N_{pm}=6)$ QLS on Community graphs.

The tradeoffs between $N_s$ and $N_{pm}$ can have significant impacts on real-world implementations which must consider the limited qubit count of the quantum computer executing the QLS ansatz. Similarly, the selection of the $(N_s, N_{pm})$ parameters can impact the tradeoff between quantum and classical computational resources: increasing $N_s$ and $N_{pm}$ means more of the graph can be explored at once using wider quantum circuits but requires optimizing over more parameters, on the other hand, smaller $N_s, N_{pm}$ results in quantum circuits with fewer qubits and variational parameters but requires more iterations to fully explore the graph.

Additionally, for equal values of $N_{pm}$, we see in Fig.~\ref{fig:convergence} that increasing the neighborhood size can also improve the local search runtime by providing more freedom for the application of partial mixers during the neighborhood ansatz construction. In Fig.~\ref{fig:convergence} this appears in the separation between same colored plots (indicating equal $N_{pm}$) with different sized neighborhoods (indicated by marker style).

In Fig.~\ref{fig:allbars} we compare the average overall performance of the different algorithms. Across all graph types and sizes, QLS is able to find the largest independent sets. However, this does come at the cost of extra local search iterations. In Fig.~\ref{fig:convergence} the CLS algorithm at each neighborhood size terminates after just a few iterations, this is because the local neighborhood search step of CLS is performed over the entire neighborhood during each iteration. Fig.~\ref{fig:circuitsizes} shows that the size of the neighborhoods that CLS optimizes over are on average much larger than those considered by QLS. This difference between the two local search strategies may account for the difference in their performance. The QLS approach is more conservative in the sense that while many qubits may participate in the local neighborhood optimization, many of those act only as control qubits, and only a subset of the qubits will be acted on by the $R_x(\beta)$ rotations that compose the partial mixers and therefore may be added to the independent set. Overall, this strategy requires more iterations of local search but results in larger independent sets when compared to CLS.

\section{Conclusion and Future Directions}\label{sec:conclusions}
While this work has focused on MIS, the QLS approach with the QAO-Ansatz can be applied to many more constrained combinatorial optimization problems.
The non-commutativity of the components of the mixing unitary within the QAO-Ansatz can be exploited by QLS for many of these problems.
Additionally, while we have restricted our simulations to 100 node graphs, this technique is easily scalable to much larger problem sizes. Additionally, more efficient circuit simulation techniques will allow us to increase the size of the local neighborhoods considered during each iteration of QLS.

In this work, we evaluated the performance of QLS against greedy random search, Boppana-Halld\'{o}rsson, local search utilizing Boppana-Halld\'{o}rsson, and QAOA+ which are all efficient heuristic algorithms, but do not promise the best solutions that are obtainable via other available classical techniques. We demonstrated the applicability of QLS to large problem sizes, but there are still many improvements that can be made to the neighborhood solution search and neighborhood update steps by drawing upon the extensive work that has been done in the classical community on local search\cite{andrade2012fast, krivelevich2019greedy, wu2015review, aarts2003local}. One advantage that QLS maintains over classical approaches to local search is its ability to exploit quantum entanglement to search the solution landscape within a neighborhood all-at-once instead of one-by-one.

Finally, implementations of the QLS algorithm on real quantum hardware will be critical for studying the impacts of noise and compilation on the algorithm's performance. The multi-qubit gates required to implement this algorithm can be quite expensive when decomposed into the single- and two-qubit gates required by current superconducting and trapped-ion architectures~\cite{vivek2009toffoli, he2017decompositions}. Using well-known decompositions, an $n$-controlled X-rotation requires $\mathcal{O}(28n)$ CNOTs~\cite{barenco1995elementary}. However, emerging quantum computer architectures, such as neutral atoms~\cite{saffman2019quantum}, are especially promising because they support the ability to natively implement multi-qubit operations with $n>2$ controls. Recent experimental efforts have also been working toward support for larger multi-qubit operations in superconducting~\cite{baekkegaard2019realization} and trapped-ion~\cite{katz2022nbody} architectures. More efficient decompositions are also possible by moving beyond two-level qubits and representing information with higher-dimensional systems such as qutrits. For example, an $n$-controlled unitary may be implemented using qutrits and $\mathcal{O}(3n)$ two-body entangling gates~\cite{gokhale2019asymptotic}.

Local search methods are an effective means of expanding the reach of current quantum systems towards larger and larger problem sizes. Here we have demonstrated the effectiveness of combining techniques from classical local search with quantum approximate optimization for the MIS problem. Progress in both quantum algorithms and hardware is proceeding rapidly, and synthesizing these advancements into a practically useful application remains a challenging open problem.

\section*{Acknowledgments}
 T.T. is supported in part by EPiQC, an NSF Expedition in Computing, under grant CCF-1730082. Z.S. and M.S. are supported by the National Science Foundation under Award No. 2037984, and by the US Department of Energy (DOE) Office of Science Advanced Scientific Computing Research (ASCR) Accelerated Research in Quantum Computing (ARQC).
 
 The submitted manuscript has been created by UChicago Argonne, LLC, Operator of Argonne National Laboratory (``Argonne”). Argonne, a U.S. Department of Energy Office of Science laboratory, is operated under Contract No. DE-AC02-06CH11357. The U.S. Government retains for itself, and others acting on its behalf, a paid-up nonexclusive, irrevocable worldwide license in said article to reproduce, prepare derivative works, distribute copies to the public, and perform publicly and display publicly, by or on behalf of the Government. The Department of Energy will provide public access to these results of federally sponsored research in accordance with the DOE Public Access Plan (\url{http://energy.gov/downloads/doe-public-access-plan}).

\textbf{Conflict of Interest}:
The authors declare that they have no confict of
interest.

\bibliographystyle{unsrt}
\bibliography{qls-quantum-submission}

\end{document}